# Solving $k$-Set Agreement with Stable Skeleton Graphs

Martin Biely*, Peter Robinson‡, and Ulrich Schmid†

* EPFL, Switzerland, biely@ecs.tuwien.ac.at
† ECS Group, Technische Universität Wien, Austria, s@ecs.tuwien.ac.at
‡ Division of Mathematical Sciences, Nanyang Technological University, Singapore, peter.robinson@ntu.edu.sg

*Abstract*—In this paper[1] we consider the $k$-set agreement problem in distributed message-passing systems using a round-based approach: Both synchrony of communication and failures are captured just by means of the messages that arrive within a round, resulting in round-by-round communication graphs that can be characterized by simple communication predicates. We introduce the weak communication predicate $\mathcal{P}_{\text{srcs}}(k)$ and show that it is tight for $k$-set agreement, in the following sense: We (i) prove that there is no algorithm for solving $(k-1)$-set agreement in systems characterized by $\mathcal{P}_{\text{srcs}}(k)$, and (ii) present a novel distributed algorithm that achieves $k$-set agreement in runs where $\mathcal{P}_{\text{srcs}}(k)$ holds. Our algorithm uses local approximations of the stable skeleton graph, which reflects the underlying perpetual synchrony of a run. We prove that this approximation is correct in all runs, regardless of the communication predicate, and show that graph-theoretic properties of the stable skeleton graph can be used to solve $k$-set agreement if $\mathcal{P}_{\text{srcs}}(k)$ holds.

## I. INTRODUCTION

The quest of finding minimal synchrony requirements for circumventing the impossibility of distributed agreement problems like consensus [9] has always been a very active research topic in distributed computing. Since the exact solvability border of consensus has been researched exhaustively, see e.g., [2], [6], [12], the attention has shifted to weaker agreement problems, in particular, $k$-set agreement [1], [11], [14], which allows the processes in a distributed system to agree on at most $k$ different values. For $k > 1$, the problem itself is possibly not as interesting as consensus ($k = 1$) from a practical point of view, except for partitionable systems that need to reach consensus in every partition. In any case, $k$-set agreement is highly relevant from a theoretical perspective, as it allows to study what level of agreement can be achieved in a fault-tolerant distributed system. This question is definitely relevant in practice, e.g., for name-space reduction (renaming) and similar problems.

One way to model synchrony requirements is through the use of round models. Round-based distributed algorithms execute in a sequence of communication-closed rounds, which consist of message exchanges and processing steps. The classic partially synchronous models of Dwork et. al. [7] were probably the first to allow some messages not to arrive within a round due to asynchrony (i.e., non-timeliness), rather than solely due to failures. The seminal work by Santoro and Widmayer [15], [16] unified the treatment of asynchrony and failures by considering synchronous processes that only suffer from "end-to-end communication failures". This idea also underlies the *Round-by-Round* failure detector (RRFD) approach by Gafni [10], which assumes a local RRFD that tells whether a process shall wait for a round message from some other process or not. The actual reason why a receiver process does not get a message from the sender process is considered irrelevant here. The Heard-Of (HO) model [3], [4] integrates this unified treatment of failures and asynchrony of [15], [16] with a flexible way of describing guarantees about communication. The basic entity of this model are communication-closed rounds and HO predicates, which specify conditions on the collection of heard-of sets: For each round $r$ and process $p$, $HO(p,r)$ denotes the set of processes that $p$ hears of (i.e., receives a message from) in round $r$.

In this paper, we will use properties of communication graphs for studying $k$-set agreement in message passing systems with very weak synchrony requirements. In $k$-set agreement, correct processes must output a single value based on values proposed locally, with no more than $k$ different values being output system-wide.

*Detailed contributions:* We introduce an algorithm for $k$-set agreement, which exploits a natural correspondence between communication predicates and round-by-round "timely communication" graphs $\mathcal{G}^r$ in a run; $\mathcal{G}^r$ contains an edge $(q \to p)$ when process $p$ *hears of* $q$ in round $r$. Our algorithm incorporates a generic method for approximating the *stable skeleton* $\mathcal{G}^{\cap\infty}$, which is the intersection of all $\mathcal{G}^r$ and reflects the underlying perpetual synchrony of a run. We also introduce the class of communication predicates $\mathcal{P}_{\text{srcs}}(k)$, which guarantees that at least two processes in every subset of $k+1$ processes hear from a common process, in every round. Using the graph-theoretic properties of $\mathcal{G}^{\cap\infty}$ guaranteed by the predicate $\mathcal{P}_{\text{srcs}}(k)$, we show that our algorithm solves $k$-set agreement in all runs where $\mathcal{P}_{\text{srcs}}(k)$ holds. Moreover, we also show that $\mathcal{P}_{\text{srcs}}(k)$ is "tight" for $k$-set agreement, as it is too weak for solving $k-1$-set agreement.

[1] Peter Robinson has been supported by the Austrian Science Foundation (FWF) project P20529 and Nanyang Technological University grant M58110000.

## II. COMPUTING MODEL AND PROBLEM DEFINITION

We consider distributed computations of a set of processes $\Pi$ communicating by message passing. Moreover, we consider that the computation is organized in an infinite sequence of communication-closed [8] rounds; that is, any message sent in a round can be received only in that round. As in the models of Gafni[10] and Charron-Bost and Schiper [4], we will express assumptions about the synchrony and the reliability of communication in a system by a predicate that characterizes the set of edges in the communication graph of each round. Intuitively speaking, there is an edge from process $p$ to $q$ in the communication graph of round $r$ is $q$ received $p$'s round $r$ message. We will in fact name a system by its predicate, that is, in a system $\mathcal{P}$ the collections of communication graphs of each run of an algorithm in that system will must fulfill predicate $\mathcal{P}$.

We now formally define computations in our round model. As in the aforementioned models, an algorithm is composed of two functions: The sending function determines, for each process $p$ and round $r > 0$, the message $p$ broadcasts in round $r$ based on the $p$'s state at the beginning of round $r$. The transition function determines, for each $p$ and round $r$ and the vector of messages received in $r$, the state at the end of round $r$, i.e., at the beginning of round $r+1$. Clearly, a *run* of an algorithm is completely determined by the initial states of the processes and the sequence of communication graphs.

For each round $r$, we denote the *communication graph* by $\mathcal{G}^r = \langle V, E^r \rangle$, where each node of the set $V$ is associated with one process from $\Pi$, and where $E^r$ is the set of directed *timely edges* for round $r$. There is an edge from $p$ to $q$, denoted as $(p \to q)$, if and only if $q$ receives $p$'s round $r$ message (in round $r$).[2] To simplify the presentation, we will denote a process and the associated node in the communication graph by the same symbols. However, as we differentiate between $V$ and $\Pi$, we will always be able to resolve possible ambiguities by stating from which set a node or process is taken. We will write $p \in \mathcal{G}^r$ and $(p \to q) \in \mathcal{G}^r$ instead of $p \in V$ resp. $(p \to q) \in E^r$.

We are primarily interested in the round $r$ skeleton $\mathcal{G}^{\cap r}$ of $\mathcal{G}^r$, which we define as the subgraph consisting of the edges that have been timely in all rounds up to round $r$. Formally, $\mathcal{G}^{\cap r} := \langle V, E^{\cap r} \rangle$ where $E^{\cap r} := \bigcap_{0 < r' \leqslant r} E^{r'}$. The crucial property of $E^{\cap r}$ is that once an edge is untimely in some round $r$, it cannot be in $\mathcal{G}^{\cap r'}$, for any $r' \geqslant r$. That is, $\forall r > 0 \colon E^{\cap r} \supseteq E^{\cap r+1}$, which implies the subgraph relation

$$\forall r > 0 \colon \mathcal{G}^{\cap r} \supseteq \mathcal{G}^{\cap r+1}. \tag{1}$$

We are particularly interested in the *stable skeleton of a run*, which we define as the intersection[3] over all rounds, i.e.,

$$\mathcal{G}^{\cap \infty} := \bigcap_{r \in \mathbb{N}^+} \mathcal{G}^{\cap r}. \tag{2}$$

Considering that a run $\alpha$ consists of infinitely many rounds, whereas our system consists of only a finite number of processes, it follows that the number of possible distinct stable skeletons must also be finite. Consequently, the subgraph property (1) implies that there is some round $r_{\text{ST}}$ when $\mathcal{G}^{\cap \infty}$ has *stabilized*, i.e., $\forall r \geqslant r_{\text{ST}} \colon \mathcal{G}^{\cap r} = \mathcal{G}^{\cap \infty}$.

As mentioned in the introduction, our algorithm will solve $k$-set agreement by approximating the stable skeleton of a run. The first step in this effort is to use the locally available information about the communication graph, which is captured by the notion of timely neighbourhoods. The *timely neighborhood of* $p$, denoted as $PT(p, r)$, is the set of processes that process $p$ has perceived as *perpetually timely* until round $r$. In other words, $p$ has received a message from every process in $PT(p, r)$ in every round up to and including $r$, i.e., $PT(p, r) := \{q \mid (q \to p) \in \mathcal{G}^{\cap r}\}$. Analogously to (1) and (2), we have

$$PT(p, r) \supseteq PT(p, r+1) \tag{3}$$

and define

$$PT(p) := \bigcap_{r > 0} PT(p, r). \tag{4}$$

We will make heavy use of the standard graph-theoretic notion of a *strongly connected component* of $\mathcal{G}^{\cap r}$. Note that we implicitly assume that strongly connected components are always nonempty and maximal. We use the superscript notation $\mathcal{C}^r$ when talking about a *strongly connected component of* $\mathcal{G}^{\cap r}$. Moreover, we write $\mathcal{C}^r_p$ to denote the (unique) strongly connected component of $\mathcal{G}^{\cap r}$ that contains process $p$ in round $r$. The strongly connected component $\mathcal{C}^\infty_p \subseteq \mathcal{G}^{\cap \infty}$ that contains $p$ in a run is defined analogously to (2) as

$$\mathcal{C}^\infty_p := \bigcap_{r > 0} \mathcal{C}^r_p.$$

Note that when $p$ and $q$ are strongly connected in $\mathcal{G}^{\cap r}$, then they are also strongly connected in all $\mathcal{G}^{\cap r'}$, for $0 < r' \leqslant r$. From property (1) of $\mathcal{G}^{\cap r}$, we immediately have

$$\forall r > 0 \colon \mathcal{C}^r_p \supseteq \mathcal{C}^{r+1}_p. \tag{5}$$

We will also use *directed paths* in $\mathcal{G}^{\cap r}$, where we assume that all nodes on a path are distinct.

Let $\mathcal{C}^r \subseteq \mathcal{G}^{\cap r}$ be a strongly connected component. If $\mathcal{C}^r$ has no incoming edges from any $q \in \mathcal{G}^{\cap r} \setminus \mathcal{C}^r$, we say $\mathcal{C}^r$ is a *root component in round* $r$. Formally,

$$\forall p \in \mathcal{C}^r \; \forall q \in \mathcal{G}^{\cap r} \colon (q \to p) \in \mathcal{G}^{\cap r} \Rightarrow q \in \mathcal{C}^r.$$

Figure 1b shows a graph with 2 root components $\{p_3, p_4, p_5\}$ and $\{p_1, p_2\}$.

Regarding the relation to the existing round-by-round models, we shortly recall what their predicates are based on: In the Heard-Of model [4], for each round $r$ and each process $p$, the set $HO(p, r)$ contains those processes that $p$ hears from, i.e., receives a message from, in round $r$. In the case of the Round-by-Round Fault Detectors [10], the output of $p$'s fault detector in round $r$ is referred to by $D(p, r)$. In each round $r$, process $p$ waits until it receives a message from every process that is not contained in $D(p, r)$. While it is possible that $p$ also receives a round $r$ message from a process in $D(p, r)$, we will consider that this is never the case. From this it is evident that we have the following correspondence between our skeleton graphs and the HO/RbR model:

$$(p \to q) \in E^{\cap r} \iff \begin{cases} \forall r' \leqslant r : p \in HO(q, r') \\ \forall r' \leqslant r : p \notin D(q, r') \end{cases} \quad (6)$$

---

[2]Since we consider communication-closed rounds, a message sent in round $r$ cannot be received in any later round.
[3]For simplicity, we set $G \cap G' := \langle V \cap V', E \cap E' \rangle$.

Thus a process can determine its timely neighbourhood in the two models as follows:

$$PT(p, r) = \begin{cases} \bigcap_{0 < r' \leqslant r} HO(p, r') \\ \Pi \setminus \left( \bigcup_{0 < r' \leqslant r} D(p, r') \right) \end{cases} \quad (7)$$

As in the HO-model, we model a crashed processes by an "internally correct" process that no other process receives messages from after it has crashed [4, Sec. 2.2]. This modelling allows us to require that all processes decide. For a more detailed discussion on the relation between models where crashed processes actually stop and the HO-model, we refer to [13].

*A. $k$-Set Agreement*

The *$k$-set agreement* problem was introduced in [5]. Every process $p$ starts with a proposal value $v$ and must eventually and irrevocably decide on some value adhering to the following three constraints:

$k$-Agreement:    Processes must decide on at most $k$ different values.
Validity:    If a process decides on $v$, then $v$ was proposed by some process.
Termination:    Every process must eventually decide.

Note that the $k$-set agreement problem was shown to be impossible in the asynchronous system model (see [1], [11], [14]) if $f \geqslant k$ processes can crash. Recalling the correspondence between crashed processes and process that no one hears of, it is not surprising that this impossibility also holds for the system $\mathcal{P}_{\text{true}} :: \text{TRUE}$, where *all* runs are admissible.

## III. A Tight Communication Predicate for $k$-Set Agreement

In this section, we introduce a predicate that, together with Algorithm 1 in Section IV, is sufficient for solving $k$-set agreement.

For a run $\alpha$, predicate $\mathcal{P}_{\text{srcs}}(k)$ requires that in every set $S$ of $k + 1$ processes, there are two processes $q, q'$ that receive timely messages from the same common process $p$, in every round. We say that $p$ is a *2-source* and $q, q'$ are *timely receivers of $p$* in $\alpha$.

$$\mathcal{P}_{\text{src}}(p, S) :: \exists q, q' \in S, q \neq q' : p \in (PT(q) \cap PT(q'))$$
$$\mathcal{P}_{\text{srcs}}(k) :: \forall S, |S| = k + 1 \; \exists p \in \Pi : \mathcal{P}_{\text{src}}(p, S) \quad (8)$$

Note that $p$ is not required to be distinct from $q$ and $q'$: $\mathcal{P}_{\text{srcs}}(k)$ still holds if $p = q$, i.e., $p$ always perceives itself in a timely fashion. Regarding communication graphs, this predicate ensures that any induced sub-graph $S$ of $\mathcal{G}^{\cap \infty}$ with $k + 1$ nodes contains distinct nodes $q$ and $q'$, such that, for some node $p$, edges $(p \to q)$ and $(p \to q')$ exist (one of which may be a self-loop). Figure 1b shows the stable skeleton graph in a run where $\mathcal{P}_{\text{srcs}}(k)$ holds for $k = 3$.

At a first glance, it might appear that the perpetual nature of $\mathcal{P}_{\text{srcs}}(k)$ is an unnecessarily strong restriction. To see why some (possibly weak) perpetual synchrony is necessary, consider the predicate $\Diamond \mathcal{P}_{\text{srcs}}(k)$ that satisfies (8) just eventually, and suppose that there is an algorithm $A$ that solves $k$-set agreement in system $\Diamond \mathcal{P}_{\text{srcs}}(k)$. Due to its "eventual" nature, $\Diamond \mathcal{P}_{\text{srcs}}(k)$ allows runs where *every* process forms a root component by itself, i.e., hears from no other process, for a finite number of rounds. Moreover, for any $k$, the (infinite) run, where a *single* process forms a root component forever and thus has to decide on its own input value, is admissible. Using a simple indistinguishability argument, it is easy to show that processes decide on $n$ different values.

The following result will be instrumental in Section IV, where we show how to solve $k$-set agreement with $\mathcal{P}_{\text{srcs}}(k)$. Note that Theorem 1 is independent of the algorithm employed.

*Theorem 1:* There are at most $k$ root components in any run that is admissible in system $\mathcal{P}_{\text{srcs}}(k)$.

*Proof:* Assume by contradiction that there is a run $\alpha$ of some algorithm $A$ that is admissible in system $\mathcal{P}_{\text{srcs}}(k)$, where there is a set of $\ell \geqslant k + 1$ disjoint root components $R = \{\mathcal{C}_{p_1}^\infty, \ldots, \mathcal{C}_{p_\ell}^\infty\}$ containing processes $p_1, \ldots, p_{k+1}, \ldots, p_\ell$. Let $r$ be the round where every strongly connected root component $\mathcal{C}_{p_i}^\infty \in R$ has stabilized, i.e., $\forall i : \mathcal{C}_{p_i}^r = \mathcal{C}_{p_i}^\infty$. That is, any two distinct root components in $R$ must already be disjoint from round $r$ on. Since $\alpha$ satisfies $\mathcal{P}_{\text{srcs}}(k)$ and $\ell \geqslant k + 1$, there must be a 2-source $p$ such that, for two distinct processes $p_i, p_j \in \{p_1, \ldots, p_{k+1}\}$, it holds that $p \in (PT(p_i) \cap PT(p_j))$. By (6), it follows that the edges $e_i = (p \to p_i)$ and $e_j = (p \to p_j)$ are in $\mathcal{G}^{\cap r}$. Considering that $\mathcal{C}_{p_i}^r$ and $\mathcal{C}_{p_j}^r$ are root components by assumption, i.e., do not have incoming edges, it must be that $e_i \in \mathcal{C}_{p_i}^r$ and $e_j \in \mathcal{C}_{p_j}^r$, and therefore $p \in \mathcal{C}_{p_i}^r \cap \mathcal{C}_{p_j}^r$. This, however, contradicts the fact that $\mathcal{C}_{p_i}^r$ and $\mathcal{C}_{p_j}^r$ are disjoint, which completes our proof. ∎

*A. Impossibility of $(k-1)$-Set Agreement*

We will now show that $\mathcal{P}_{\text{srcs}}(k)$ does *not* allow to solve $(k-1)$-set agreement. More specifically, we will prove this by assuming the existence of such an algorithm $A$, and then construct a run fulfilling $\mathcal{P}_{\text{srcs}}(k)$ where processes decide on $k$ (instead of $k - 1$) different values.

*Theorem 2:* Consider any $k$ such that $1 < k < n$. There is no algorithm $A$ that solves $(k-1)$-set agreement in system $\mathcal{P}_{\text{srcs}}(k)$.

*Proof:* Assume for the sake of a contradiction that such an algorithm $A$ exists. Suppose that all processes start with pairwise distinct input values. Consider the run $\alpha$ and a fixed set $L$ of $k - 1$ processes that only hear from themselves, formally speaking, $\forall p \in L : PT(p) = \{p\}$. Moreover, there is one process $s$ such that every process not in $L$ only hears from itself and $s$, i.e.,

$$\forall p \in \Pi \setminus L \colon PT(p) = \{p, s\}.$$

Since, by *validity* and *termination*, processes eventually have to decide on some input value and processes in $L \cup \{s\}$ cannot learn any other process' input value, they have to decide on their own value. Thus, we have $k$ different decision values, as we have assumed a unique input value for each process, and therefore a violation of $(k-1)$-*agreement*.

What remains to be shown is that this run $\alpha$ actually fulfills $\mathcal{P}_{\text{srcs}}(k)$. Recall equation (8), i.e., the definition of $\mathcal{P}_{\text{srcs}}$, and consider for any set $S$ of size $k+1$ the set $P = S \setminus L$. Since $|S \setminus L| \geqslant 2$, the set $P$ contains at least two distinct processes that permanently hear from $s$ (one of which may be $s$). That is, process $s$ is the required 2-source for any set $S$ of $k+1$ processes. ∎

## IV. APPROXIMATING THE STABLE SKELETON GRAPH AND SOLVING $k$-SET AGREEMENT

In this section, we present and analyze an algorithm that solves $k$-set agreement with predicate $\mathcal{P}_{\text{srcs}}(k)$. Algorithm 1 employs a generic approximation of the stable skeleton graph of the run, which works as follows:

First, every process $p$ keeps track of the processes it has perceived as timely until round $r$ in the set $PT_p$, updated in Line 9. Lemma 3 will show that $PT_p$ satisfies the definition of $PT(p, r)$, for all rounds $r$. In addition, every process $p$ locally maintains an approximation graph $G_p$ of the stable skeleton, denoted $G_p^r$ for round $r$, which is broadcast in every round. If a process $q$ receives such a graph $G_p^r$ from some process $p$ in its timely neighborhood $PT(q, r)$, it adds the information contained in $G_p^r$ to its own local approximation $G_q^r$. Note that, in contrast to the stable skeleton graph $\mathcal{G}^{\cap r}$, the approximation graph $G_p$ is actually a *weighted* directed graph. The edge labels of $G_p$ correspond to the round number when a particular edge was added by some process, i.e., the edge $(q' \stackrel{r}{\to} q)$ is in $G_p$ if, and only if, $q' \in PT(q, r)$ (cf. Lemma 3(b)). To prevent outdated information from remaining in the approximation graph permanently, every process $p$ purges all edges in $G_p^r$ that were initially added more than $n-1$ rounds ago. Figures 1c-1h show this approximation mechanism at work.

For $k$-set agreement, process $p$ only considers proposal values for its estimated decision value $x_p$ that were sent by processes in its current timely neighborhood, i.e., in $PT_p$. This ensures that $p$ and $q$ will have a common estimated decision value $x_p = x_q$ in round $n$, if they are in the same strongly connected component (cf. Lemma 14). To determine when to terminate, $p$ analyzes its approximation graph in every round $r \geqslant n$ and decides if $G_p^r$ is a strongly connected graph.

Why is this decision safe with respect to the agreement property? Using our graph approximation results, we will show in Lemma 15 that any strongly connected approximation graph contains at least one root component in the stable skeleton graph. Furthermore, if two processes decide on different

---

**Algorithm 1** *Approximating the stable skeleton graph and solving $k$-set agreement with $\mathcal{P}_{\text{srcs}}(k)$*

**Variables and Initialization:**
1: $PT_p \in 2^\Pi$ initially $\Pi$
2: $x_p \in \mathbb{N}$ initially $v_p$ // Estimated decision value
3: $G_p := \langle V_p, E_p \rangle$ initially $\langle \{p\}, \emptyset \rangle$ // weighted digraph
4: $decided_p \in \{0, 1\}$ initially 0 // is 1 iff $p$ has decided

**Round $r$:** sending function $S_p^r$:
5: **if** $decided_p = 1$ **then**
6:     send $(decide, x_p, G_p)$ to all processes
7: **else**
8:     send $(prop, x_p, G_p)$ to all processes

**Round $r$:** transition function $T_p^r$:
9: update $PT_p$
10: **if** received $(decide, x_q, \_)$ from $q \in PT_p$ and $decided_p = 0$ **then**
11:     $x_p \leftarrow x_q$
12:     decide on $x_p$
13:     $decided_p \leftarrow 1$

14: // Approximate stable skeleton graph:
15: $G_p \leftarrow \langle \{p\}, \emptyset \rangle$
16: **for** $q \in PT_p$ **do**
17:     add directed edge $(q \stackrel{r}{\to} p)$ to $E_p$
18:     $V_p \leftarrow V_p \cup V_q$
19: **for** every pair of nodes $(p_i, p_j) \in V_p \times V_p$ **do**
20:     $R_{i,j} \leftarrow \{r_e \mid \exists q \in PT_p \colon (p_i \stackrel{r_e}{\to} p_j) \in E_q\}$
21:     **if** $R_{i,j} \neq \emptyset$ **then**
22:         $r_{max} \leftarrow \max(R_{i,j})$
23:         $E_p \leftarrow E_p \cup \{(p_i \stackrel{r_{max}}{\to} p_j)\}$
24: discard all $(p_i \stackrel{r_e}{\to} p_j)$ from $E_p$ where $r_e \leqslant r - n$
25: discard $p_i \neq p$ from $V_p$ if $p$ is unreachable from $p_i$ in $G_p$

26: **if** $decided_p = 0$ **then**
27:     $x_p \leftarrow \min \{x_q \mid q \in PT_p\}$
28:     **if** $r \geqslant n$ and $G_p$ is strongly connected **then**
29:         decide on $x_p$
30:         $decided_p \leftarrow 1$

---

values, it follows that their approximated graphs in the rounds of their respective decision are disjoint. Since Theorem 1 confirms that there are at most $k$ root components in any run where $\mathcal{P}_{\text{srcs}}(k)$ holds, there can be in fact at most $k$ different decision values.

### A. Approximation of the Stable Skeleton Graph

Throughout our analysis, we denote the value of variable $var$ of process $p$ at the end of round $r$ as $var_p^r$. When we use the subgraph relation ($\subseteq$) between graphs $\mathcal{C}_p^r$ and $G_p^r$, we mean the standard subgraph relation between $\mathcal{C}_p^r$ and the *unweighted* version of $G_p^r$. We first state some obvious facts that follow directly from the code of the algorithm:

*Observation 1:* For any round $r > 0$ it holds that $p \in G_p^r$ and that no edge $(q' \stackrel{s}{\to} q) \in G_p^r$ has $s \leqslant r - n$.

Note that, after the initial assignment, $p$ only updates variable $PT_p$ in Line 9, which is equivalent to (7). From this and the inspection of Lines 15 and 17, Lemma 3 follows immediately:

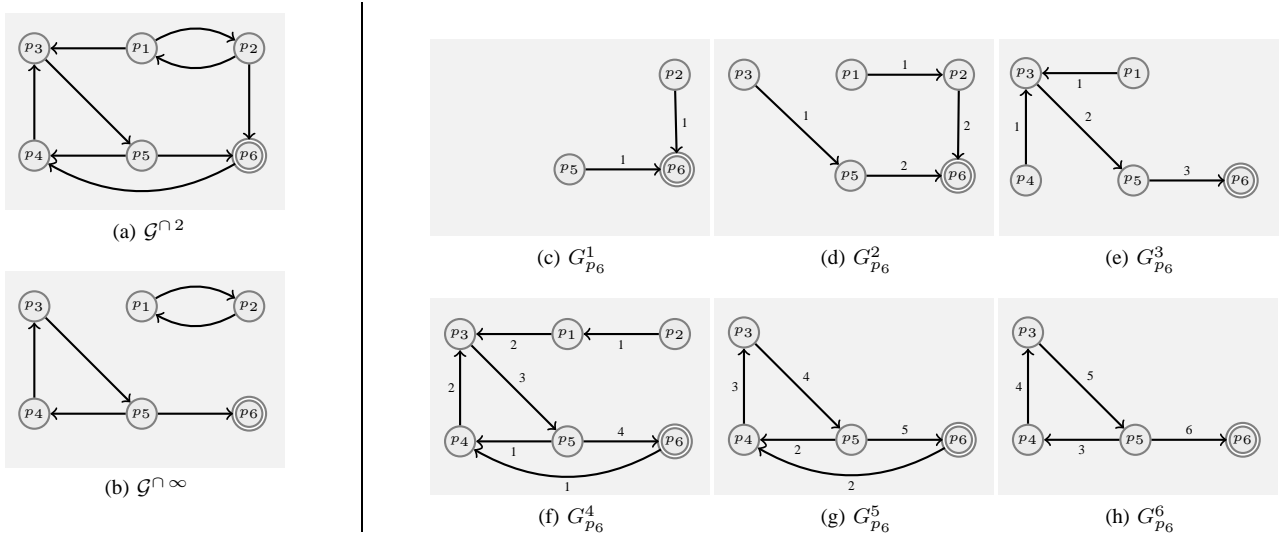

Fig. 1: A system of 6 processes where $\mathcal{P}_{\mathrm{srcs}}(3)$ holds. The stable skeleton graph for round 2 is depicted in Figure 1a; 1b shows the stable skeleton graph for the entire run. For simplicity, we omit self-loops, i.e., $\forall p_i \colon p_i \in PT(p_i)$. Figures 1c-1h show process $p_6$'s approximation of $\mathcal{G}^{\cap \infty}$ during rounds 1 to 6.

*Lemma 3:* It holds that $q \in PT(p, r)$ if, and only if, all of the following are true:
(a) $q \in PT_p^r$,
(b) $p$ adds a directed edge $q \xrightarrow{r} p$ to $G_p^r$ by executing Line 17 in round $r$, and
(c) for any $r' \neq r$, there is no other edge $q \xrightarrow{r'} p$ in $G_p^r$.

The following lemma shows that the approximation graph $G_{p_{\ell+1}}$ accurately reflects the timely neighborhood of a process. That is, if $p_1$ is connected to $p_{\ell+1}$ through a path of length $\ell$, then $p_{\ell+1}$ will add the timely neighborhood information of $p_1$ to its approximated graph by round $\ell$.

*Lemma 4:* Suppose that there exists a directed path

$$\Gamma = (p_1 \to \ldots \to p_{\ell+1})$$

in $\mathcal{G}^{\cap r}$ for round $r \geqslant n$, where $\Gamma$ has length $\ell \leqslant n-1$. Then, $\forall q \in PT(p_1, r-\ell)$ it holds that
(a) edge $(q \xrightarrow{r_q} p_1)$ is in $G_{p_{\ell+1}}^r$ where $r \geqslant r_q \geqslant r-\ell$, and
(b) $G_{p_{\ell+1}}^r$ contains no other edges from $q$ to $p_1$.

*Proof:* Consider an arbitrary $q \in PT(p_1, r-\ell)$. The proof proceeds by induction over the edges of path $\Gamma$ indexed by $k$. That is, we show that for all $k$, with $0 \leqslant k \leqslant \ell$, it holds that there is an edge $e = (q \xrightarrow{r_k} p_1)$ in $G_{p_{1+k}}^{r-\ell+k}$ where $r-\ell+k \geqslant r_k \geqslant r-\ell$.

For the base case ($k = 0$), we have to show that the edge $e$ is in $G_{p_1}^{r-\ell}$, but this already follows from $q \in PT(p_1, r-\ell)$, by Lemma 3.

For the induction step, we assume that the statement holds for some $k < \ell$ and then show that it holds for $k+1$ as well. In round $r-\ell+(k+1)$ process $p_{1+k}$ broadcasts its current graph estimate, i.e., $G_{p_{1+k}}^{r-\ell+k}$ to all. We know that $p_{1+(k+1)}$ will receive this message since $(p_{1+k} \to p_{1+(k+1)})$ is in the path $\Gamma \subseteq \mathcal{G}^{\cap r}$, which means that

$$p_{1+k} \in PT(p_{1+(k+1)}, r-\ell+(k+1)).$$

By the induction hypothesis, the edge $(q \xrightarrow{r_k} p_1)$ is in $G_{p_{1+k}}^{r-\ell+k}$ and therefore will be among the edges that $p_{1+(k+1)}$ considers in Line 20. This in turn implies that $p_{1+(k+1)}$ will add an edge $q \xrightarrow{r_{k+1}} p_1$ to its graph $G_{p_{1+(k+1)}}^{r-\ell+(k+1)}$ in Line 23, whereby $r_{k+1}$ is calculated in Line 22 such that $r_{k+1} \geqslant r_k$. Moreover, by induction hypothesis we have $r_k \geqslant r-\ell > r-n$, which ensures that the edge will not be discarded in Line 24. Since the code following the for-loop in Line 19 is executed exactly once for every edge, no other edge $q \xrightarrow{r'} p_1$ is added to $G_{p_{1+(k+1)}}^{r-\ell+(k+1)}$. This completes the proof our lemma. ∎

The next lemma shows that the approximation graph of correctly (over)estimates the strongly connected component from round $n$ on:

*Lemma 5:* Let $r \geqslant n$ and consider the strongly connected component $\mathcal{C}_p^r$ containing $p$ in $\mathcal{G}^{\cap r}$. Then, it holds that $G_p^r \supseteq \mathcal{C}_p^r$.

*Proof:* Consider any edge $(q' \to q) \in \mathcal{C}_p^r$. Since $\mathcal{C}_p^r$ is strongly connected, there is a directed path between any pair of processes in $\mathcal{C}_p^r$, in particular there is a path of length $\ell \leqslant n-1$ from $q$ to $p$. By the definition of $\mathcal{C}_p^r$ we know that $q$ always perceives $q'$ as timely in all rounds up to round $r$, which means that $q' \in PT(q, r-\ell)$. Then, by applying Lemma 4, we get that the edge $(q' \xrightarrow{r'} q)$ is in $G_p^r$, for some $r' \geqslant r-\ell$, which shows that $\mathcal{C}_p^r$ is a subgraph of $G_p^r$. ∎

Lemma 3 showed that the timely neighborhood is eventually in the approximated graph. We now show that our approximation contains only valid information:

*Lemma 6:* Let $r \geqslant 1$ and suppose that there is an edge $e = (q' \xrightarrow{s} q)$ in the approximated stable skeleton graph $G_p^r$ of process $p$. Then it holds that $q' \in PT(q, s)$.

*Proof:* Note that processes only add edges to their approximation graphs in Line 17 or in Line 23. If an edge is added via Line 23, then this edge has previously been added by another process by executing Line 17. Therefore, every edge must have been added by some process via Line 17. In case of $e$, this process can only be $q$. By Lemma 3 this happens in round $s$ and $q' \in PT(q,s)$. ∎

The following Lemma 7 is in some sense the converse result of Lemma 5, as it states that the approximated graph must approach $\mathcal{C}_p^r$ from below, if it is strongly connected:

*Lemma 7:* Let $r \geqslant 1$ and consider the strongly connected component $\mathcal{C}_p^r$. If the approximated skeleton graph $G_p^{r+n-1}$ is strongly connected, then $\mathcal{C}_p^r \supseteq G_p^{r+n-1}$.

*Proof:* Consider any edge

$$e = (q' \xrightarrow{r'} q) \in G_p^{r+n-1}.$$

By Lemma 6, we know that $q' \in PT(q,r')$. It follows by the subset property (3) that $q' \in PT(q,r)$, as Observation 1 implies

$$r' > (r+n-1) - n = r-1.$$

Therefore, there is an edge $(q' \to q)$ in $\mathcal{G}^{\cap r}$.

It follows that $G_p^{r+n-1}$ is isomorphic to a (not necessarily maximal) strongly connected component $\mathcal{S}^r$ in $\mathcal{G}^{\cap r}$. Because $\mathcal{C}_p^r$ and $\mathcal{S}^r$ both contain $p$, their intersection is nonempty, i.e., $\mathcal{C}_p^r \supseteq G_p^{r+n-1}$. ∎

As a final result about the approximated skeleton graph, we show that once the approximation $G_p$ is strongly connected in round $r \geqslant n$, it is closed w.r.t. strongly connected components. This means that $G_p$ can be partitioned into disjoint strongly connected components in $\mathcal{G}^{\cap \infty}$.

*Theorem 8:* Suppose that $R \geqslant n$. If the approximated skeleton graph $G_p^R$ is strongly connected, then it contains the strongly connected component $\mathcal{C}_q^\infty$ of every $q \in G_p^R$.

*Proof:* Consider any $q \in G_p^R$ and its strongly connected component $\mathcal{C}_q^\infty$. From (5) and Lemma 7 it follows that

$$q \in G_p^R \subseteq \mathcal{C}_p^{R-n+1} \subseteq \mathcal{C}_p^1,$$

i.e., $q \in \mathcal{C}_p^1 \cap \mathcal{C}_q^1$. Moreover, due to the well-known fact that two maximal strongly connected components in a digraph are either disjoint or equivalent, we get that $\mathcal{C}_q^1 = \mathcal{C}_p^1$.

Now suppose the theorem does not hold. Then there exists some $q' \in \mathcal{C}_q^\infty$ such that $q' \notin G_p^R$. Due to Lemma 5, $q'$ cannot be contained in $\mathcal{C}_p^R$, but due to (5), $q' \in \mathcal{C}_q^R \supseteq \mathcal{C}_q^\infty$. Therefore, $\mathcal{C}_q^R \neq \mathcal{C}_p^R$, and thus $\mathcal{C}_q^R \cap \mathcal{C}_p^R = \emptyset$. Since $G_p^R$ is strongly connected and contains $q$, it also contains a path

$$\Gamma = (q = p_\ell \to \cdots \to p_0 = p),$$

such that

$$\forall i, \ 0 \leqslant i < \ell : p_{i+1} \in PT(p_i, R-i).$$

Let $j$ be the minimal index $i$ such that $p_j \in \mathcal{C}_q^R$, and let $\Gamma_j = (p_j \to \cdots \to p_0)$ be the path remaining from $p_j$.

As both $q'$ and $p_j$ are in $\mathcal{C}_q^R$, there is a path $\Gamma'$ in $\mathcal{C}_q^R$. Let $k$ be the length of this path. Moreover, by applying Lemma 4, we get that $G_{p_j}^{R-j}$ contains the outgoing edge $e$ of $q'$ on this path, labeled with some round

$$r' \geqslant R - j - k. \quad (9)$$

But then, by the definition of $\Gamma$, it follows that when $G_p^R$ contains $p_j$ — which it does — then it must also contain $q'$, unless some process $p_i$ ($i < j$) removed $e$ from its set of edges in line 24 in round $R-i$ because $r' \leqslant R-i-n$. Since round $R$ at process $p(=p_0)$ is the latest round when this can occur, we get that $r' \leqslant R-n$, and thus, by (9),

$$R - j - k \leqslant r' \leqslant R - n, \text{ i.e., } j+k \geqslant n. \quad (10)$$

Let $\Delta$ be the subgraph obtained by concatenating paths $\Gamma'$ and $\Gamma_j$. By construction, $\Gamma_j$ and $\Gamma'$ only share node $p_j$, and thus $\Delta$ is a (simple) path and must have length $j+k \leqslant n-1$, as no path can exceed length $n-1$. This contradicts (10) and thus completes the proof that $q'$ is in $G_p^R$. The proof showing that all edges of $\mathcal{C}_q^\infty$ are in $G_p^R$ proceeds analogously, by assuming that some edge in $\mathcal{C}_q^\infty$ ending in $q'$ is not in $G_p^R$. ∎

### B. k-Set Agreement

In this section, we will show that Algorithm 1 not only approximates the stable skeleton graph, but also solves $k$-set agreement. Our previous results allow us to immediately prove the validity and the termination properties.

*Lemma 9 (Validity):* If a process decides on $v$, then $v$ was the initial value of some process.

*Proof:* Observe that the decision value $x_p$ of any process $p$ is initially set to its proposal value $v_p$, which is then broadcast. On all subsequent updates of $x_p$ in Line 27, a value $x_q$ that was sent by some process $q$ (which originated from some $v_{q'}$) is assigned, therefore *validity* holds. ∎

*Lemma 10:* Every process decides at most once in any run.

*Proof:* Observe that no process executes Line 29 and Line 12 in the same run. This is guaranteed by the fact that process $p$ cannot pass the if-conditions in Line 10 or in Line 26 after $decided_p$ is set to 1, which happens whenever $p$ decides. ∎

*Lemma 11 (Termination):* Every process decides exactly once.

*Proof:* Lemma 10 shows that every process decides at most once. We will now show that every process decides at least once. First, we will show that there is a root component in every round. Consider the strongly connected components that partition the set of nodes of the stable skeleton graph $\mathcal{G}^{\cap r}$ in some round $r$. Such a set always exists, since the strongly connected components form equivalence classes of nodes. It is well known that the contraction of the strongly connected components is a directed acyclic graph, which reveals that there is at least one node $\mathcal{C}^r$ in the contracted graph that has no incoming edges. Clearly, $\mathcal{C}^r$ satisfies the definition of a root component in $\mathcal{G}^{\cap r}$. Therefore, there is a nonempty set $R^r$ of strongly connected components all of which are root components in round $r$.

Let $r \geqslant 1$ be the earliest round where $\mathcal{G}^{\cap r}$ is stable for at least $n-1$ rounds, i.e., $\forall r' \in [r, r+n-1]: \mathcal{G}^{\cap r'} = \mathcal{G}^{\cap r}$. Note that property (1) implies that $r$ exists. Now, consider any root component $\mathcal{R}^r \in R^r$: Clearly, since every process is in exactly one strongly connected component, we have

$$\forall p \in \mathcal{R}^r : \mathcal{C}_p^r = \mathcal{R}^r = \mathcal{R}^{r+n-1} = \mathcal{C}_p^{r+n-1}. \quad (11)$$

We will now show that the approximated skeleton graph of such a process $p$ is in fact exactly the strongly connected component of $p$. Consider any $p \in \mathcal{R}^r (= \mathcal{C}_p^{r+n-1})$. First, since $(r+n-1) \geqslant n$, Lemma 5 and (11) imply that $\mathcal{R}^r \subseteq G_p^{r+n-1}$. We will now show that $\mathcal{R}^r \supseteq G_p^{r+n-1}$, which proves that these graphs are equal: Since $G_p^{r+n-1}$ is connected by construction, it is sufficient to show that every edge in $G_p^{r+n-1}$ is also in $\mathcal{R}^r$. Assume in contradiction that there is an edge $e = (q' \xrightarrow{r'} q)$ in $G_p^{r+n-1}$ such that $q \in \mathcal{R}^r$ but $q' \notin \mathcal{R}^r$; note that the other way round ($q' \in \mathcal{R}^r$ but $q \notin \mathcal{R}^r$) is impossible by construction. Using Lemma 6 we know that $q' \in PT(q, r')$, and Observation 1 implies that $r' > (r+n-1) - n = r - 1$, i.e., $r' \geqslant r$. Then, by definition, we have that $e \in \mathcal{G}^{\cap r}$, i.e., $e$ is an incoming edge of $\mathcal{R}^r$, contradicting the assumption that $\mathcal{R}^r$ is a root component. We can therefore conclude that $\mathcal{R}^r = G_p^{r+n-1}$.

By assumption, $\mathcal{R}^r$ is a root component, which tells us that $G_p^{r+n-1}$ is strongly connected, i.e., $p$ will pass the if-condition in Line 28 in round $r+n-1$ and decide. Recall the contracted stable skeleton graph of round $r+n-1$. Since every path in this graph is rooted at some node corresponding to a root component in the set $R^r$. Thus, all processes that are not in a root component will receive a decision message by round $r + 2n - 1$ and also decide, which completes our proof. ∎

In the remainder of this section we will prove that Algorithm 1 satisfies the $k$-agreement property. We will start out with some basic invariants on decision estimates.

*Observation 2 (Monotonicity):* In any run of Algorithm 1 it holds that $\forall r > 0: x_p^r \geqslant x_p^{r+1}$.

*Lemma 12:* If process $p$ does not decide in Line 12, we have that $\forall r \geqslant n - 1: x_p^r = x_p^{r+1}$.

*Proof:* Suppose that there is an $r \geqslant n - 1$ such that $p$ sets $x_p^{r+1} \leftarrow x_q$ and $x_p^r \neq x_q$. This can only occur in Line 27, if the process does not decide in Line 12. From Observation 2 and *validity* (cf. Lemma 9), we know that $p$ did not previously receive $x_q$ and that $x_q$ is the initial value of some distinct process $q$. Since processes forward their estimated decision value in every round, (3) implies that the shortest path from $q$ to $p$ (along which $x_p$ has been propagated to $p$) in $\mathcal{G}^{\cap r+1}$ has length $r + 1$. However, this is impossible as $r + 1 \geqslant n$ and the longest possible path has length $n - 1$. ∎

*Lemma 13:* Suppose that some process $p$ decides on $x_p$ in round $r$ by executing line 12. Then some process $q \neq p$ has decided on $x_p$ in round $r' < r$ by executing Line 29.

*Proof:* Every process decides either in Line 29 or in Line 12, but not both (Lemma 10). Since $p$ decided in Line 12 it must have received a $(decide, x_q, \_)$ message from some distinct process $q$. If $q$ decided in Line 29 we are done; otherwise $q$ decided in Line 12 in round $r - 1$, we can repeat the same argument for $q$. After at most $n - 1$ iterations, we arrive at some process that must have decided using Line 29. ∎

*Lemma 14:* Let $\mathcal{C}_p^n$ be the strongly connected component of process $p$ in round $n$. Then, it holds that $\forall q \in \mathcal{C}_p^n : x_q^n = x_p^n$.

*Proof:* First, observe that due to Lemma 13 and the fact that no process can pass the check in Line 28 before round $n$, no process can decide before round $n$. Therefore, processes can update their estimate values until at least round $n$.

Suppose that there are processes $p, q \in \mathcal{C}_p^n$, such that $x_p^n \neq x_q^n$. In particular we assume without loss of generality, that $x_q^n$ is minimal among all round $n$ estimation values of processes in $\mathcal{C}_p^r$, i.e., $x_p^n > x_q^n$.

Let $r_q$ be the round where $q$ first sets $x_q$ to the value $x_q^n$. By Observation 2 it follows that $q$ does not update $x_q$ anymore before round $n$. Since Algorithm 1 satisfies *validity* (Lemma 9), we know that there is some process $s$ that is the source of this value, i.e., $s$ initially proposed $x_q^r$. By the code of the algorithm we know that in round $r$ process $p$ only considers values in Line 27 that were sent by some process in $PT(p, r)$. This implies that there is a sequence of pairwise distinct processes $s = q_1, \ldots, q_\ell = q$, such that

$$\forall i, (1 \leqslant i < \ell): q_i \in PT(q_{i+1}, i). \quad (12)$$

Clearly, $r_q = \ell - 1$. Let $j \leqslant \ell$ be such that $q_j \in \mathcal{C}_p^n$ and $j$ is minimal, let $\Gamma_q$ be the path in $\mathcal{G}^{\cap 1}$ induced by the sequence $s$ up to $q_j$. Moreover, since $q_j \in \mathcal{C}_p^n$, there is a path $\Gamma_p$ in $\mathcal{C}_p^n$ from $q_j$ to $p$. Since $\mathcal{C}_p^n \subseteq \mathcal{G}^{\cap 1}$, $\Gamma_p$ is a path in $\mathcal{G}^{\cap 1}$ as well. Let $\Gamma$ be the path in $\mathcal{G}^{\cap 1}$ obtained by appending $\Gamma_p$ to $\Gamma_q$. By construction $\Gamma$ is simple, and therefore its length is bounded by $n-1$. Moreover, the initial value of $s$ was propagated along this path — over $\Gamma_q$ by construction and over $\Gamma_p$, because $x_q^n$ is minimal in $\mathcal{C}_p^n$. This leads to process $p$ assigning this value to $x_p$ in some round $r_p \leqslant n-1$, which contradicts the assumption that $x_p^n > x_q^n$. ∎

*Lemma 15 (k-Agreement):* Processes decide on at most $k$ distinct values.

*Proof:* For the sake of a contradiction, assume that there is a set of $\ell > k$ processes $D = \{p_1, \ldots, p_\ell\}$ in a run $\alpha$ where $p_i$ decides on $x_i^\infty = x_i^{r_i}$ [4] in round $r_i \geqslant n$ and $\forall p_i, p_j \in D: x_{p_i}^\infty \neq x_{p_j}^\infty$. By virtue of Lemma 13, we can assume that every $p_i$ has decided by executing Line 29. Considering that no process decides before round $n$, applying Lemma 12 yields that

$$\forall r \geqslant n \; \forall p_i, p_j \in D: x_{p_i}^r \neq x_{p_j}^r. \quad (13)$$

Note that the approximated skeleton graphs $G_{p_i}^{r_i}$ and $G_{p_j}^{r_j}$ are strongly connected in round $r_i$ resp. $r_j$, otherwise the processes could not have passed the if-condition before Line 29.

We will first show that the different decision values of $p_i$ and $p_j$ imply that their approximated skeleton graphs in rounds $r_i$ resp. $r_j$ are disjoint. Lemma 7 reveals that these skeleton graphs are contained within the respective strongly connected components of an earlier round, i.e.,

$$\mathcal{C}_{p_i}^{r_i - n + 1} \supseteq G_{p_i}^{r_i} \text{ and } \mathcal{C}_{p_j}^{r_j - n + 1} \supseteq G_{p_j}^{r_j}.$$

If these strongly connected components of $p_i$ and $p_j$ are disjoint, then so are the approximated skeleton graphs and

---
[4] Note that $x_p^\infty$ denotes $p$'s final "estimate", i.e., the actual decision value of process $p$.

we are done. Therefore, assume in contradiction that

$$I = \mathcal{C}_{p_i}^{r_i-n+1} \cap \mathcal{C}_{p_j}^{r_j-n+1} \neq \emptyset.$$

We will now prove that one of these components contains the other. Without loss of generality, suppose that $r_i \leqslant r_j$ and consider any node $p \in I \subseteq \mathcal{C}_{p_j}^{r_j-n+1}$. Clearly, $p$ is strongly connected to every node in $\mathcal{C}_{p_j}^{r_j-n+1}$. Let $\mathcal{Z}$ be the induced subgraph of $\mathcal{C}_{p_j}^{r_j-n+1}$ in the skeleton graph $\mathcal{G}^{\cap r_i-n+1}$. By the subgraph property (5) and since $r_i \leqslant r_j$, it follows that $\mathcal{Z} = \mathcal{C}_{p_j}^{r_j-n+1}$, and hence $\mathcal{Z} \cap \mathcal{C}_{p_i}^{r_i-n+1} \neq \emptyset$. By the fact that $p \in I$, we know that $p \in \mathcal{C}_{p_i}^{r_i-n+1}$. That is, in the skeleton graph $\mathcal{G}^{\cap r_i-n+1}$, process $p$ is strongly connected to all nodes in $\mathcal{C}_{p_i}^{r_i-n+1}$ and $\mathcal{Z}$. But since the strongly connected component $\mathcal{C}_{p_i}^{r_i-n+1}$ is maximal, we actually have

$$\mathcal{C}_{p_i}^{r_i-n+1} \supseteq \mathcal{Z} = \mathcal{C}_{p_j}^{r_j-n+1},$$

which means that $p_j \in \mathcal{C}_{p_i}^{r_i-n+1}$. Then, Lemma 14 readily implies that $\forall q \in \mathcal{C}_{p_i}^{r_i-n+1}$ it holds that $x_{p_i}^n = x_q^n$ and, in particular, $x_{p_i}^n = x_{p_j}^n$, which contradicts (13). We can therefore conclude that the intersection of the strongly connected components, and therefore, by Lemma 7, also the intersection of $G_p^{r_i}$ and $G_{p_j}^{r_j}$ is indeed empty, i.e.,

$$\forall p_i, p_j \in D : (G_{p_i}^{r_i} \cap G_{p_j}^{r_j}) = \emptyset. \quad (14)$$

By Theorem 8 it follows that each of the strongly connected approximated skeleton graphs $G_{p_i}^{r_i}$ can be partitioned into a set $D_i$ of strongly connected components in $\mathcal{G}^{\cap \infty}$. By Theorem 1, at most $k$ of the sets $D_i$ can contain a root component. Note that (14) implies that no strongly connected component is in two distinct sets $D_i$, $D_j$. For the sake of a contradiction, assume that (w.l.o.g.) the set $D_\ell$ corresponding to $G_{p_\ell}^{r_\ell}$ does not contain a root component. Now consider the contracted graph of $\mathcal{G}^{\cap \infty}$ where the nodes are the strongly connected components. Since the contracted graph is acyclic, it follows that there exists a path $\Gamma$ in the (non-contracted) graph $\mathcal{G}^{\cap \infty}$ that ends at process $p_\ell \in D_\ell$, and is rooted at some process $q \in \mathcal{C}_q^\infty$ where $\mathcal{C}_q^\infty$ is a root component and thus by assumption not in $D_\ell$. However, by the subgraph property (1), we know that the path $\Gamma$ is also in $\mathcal{G}^{\cap r_\ell}$. But then Lemma 4 implies that $q \in G_{p_i}^{r_i}$, and Theorem 8 shows that $\mathcal{C}_q^\infty \in D_\ell$, i.e., one of the components in $D_\ell$ in fact *is* a root component. This provides the required contradiction. ∎

*Theorem 16:* Algorithm 1 solves $k$-set agreement in system $\mathcal{P}_{\text{srcs}}(k)$.

*Proof:* Lemma 15 implies $k$-agreement. *Termination* is guaranteed by Lemma 11 and Lemma 9 shows that *validity* holds. ∎

## V. Discussion and Future Work

We have introduced the notion of communication graphs and presented an algorithm that approximates the stable skeleton of a run. The algorithm is based on exchanging local approximations of the stable skeleton, hence has a worst-case message bit complexity that is polynomially in $n$. We have also introduced a class of communication predicates $\mathcal{P}_{\text{srcs}}(k)$ and proved that using this approximation one can solve $k$-set agreement in a system that guarantees $\mathcal{P}_{\text{srcs}}(k)$. Note that the algorithm actually solves consensus in sufficiently well-behaved runs.

The one-to-one correspondence between the (at most) $k$ root components of the stable skeleton graph and distinct decision values shows that these communication graphs are a promising new tool for studying the underlying synchrony in a system. Since our algorithm yields a correct approximation atop of *any* communication predicate, part of our future work will be devoted to finding a graph-theoretic characterization of the weakest synchrony requirements for different agreement problems and further exploring the duality between communication predicates and graph-theoretic properties.

## References

[1] E. Borowsky and E. Gafni. Generalized FLP impossibility result for t-resilient asynchronous computations. In *STOC '93: Proceedings of the twenty-fifth annual ACM symposium on Theory of computing*, pages 91–100, New York, NY, USA, 1993. ACM.

[2] T. D. Chandra, V. Hadzilacos, and S. Toueg. The weakest failure detector for solving consensus. *Journal of the ACM*, 43(4):685–722, June 1996.

[3] B. Charron-Bost and A. Schiper. Improving fast Paxos: being optimistic with no overhead. In *12th IEEE Pacific Rim International Symposium on Dependable Computing (PRDC 2006)*, pages 287–295. IEEE Computer Society, 2006.

[4] B. Charron-Bost and A. Schiper. The Heard-Of model: computing in distributed systems with benign faults. *Distributed Computing*, 22(1):49–71, Apr. 2009.

[5] S. Chaudhuri. More choices allow more faults: set consensus problems in totally asynchronous systems. *Inf. Comput.*, 105(1):132–158, 1993.

[6] D. Dolev, C. Dwork, and L. Stockmeyer. On the minimal synchronism needed for distributed consensus. *Journal of the ACM*, 34(1):77–97, Jan. 1987.

[7] C. Dwork, N. Lynch, and L. Stockmeyer. Consensus in the presence of partial synchrony. *Journal of the ACM*, 35(2):288–323, Apr. 1988.

[8] T. Elrad and N. Francez. Decomposition of distributed programs into communication-closed layers. *Science of Computer Programming*, 2(3):155–173, 1982.

[9] M. J. Fischer, N. A. Lynch, and M. S. Paterson. Impossibility of distributed consensus with one faulty process. *Journal of the ACM*, 32(2):374–382, Apr. 1985.

[10] E. Gafni. Round-by-round fault detectors (extended abstract): unifying synchrony and asynchrony. In *Proceedings of the Seventeenth Annual ACM Symposium on Principles of Distributed Computing*, pages 143–152, Puerto Vallarta, Mexico, 1998. ACM Press.

[11] M. Herlihy and N. Shavit. The asynchronous computability theorem for t-resilient tasks. In *STOC '93: Proceedings of the twenty-fifth annual ACM Symposium on Theory of computing*, pages 111–120, New York, NY, USA, 1993. ACM.

[12] M. Hutle, D. Malkhi, U. Schmid, and L. Zhou. Chasing the weakest system model for implementing omega and consensus. *IEEE Transactions on Dependable and Secure Computing*, 6(4):269–281, 2009.

[13] M. Hutle and A. Schiper. Communication predicates: A high-level abstraction for coping with transient and dynamic faults. In *37th Annual IEEE/IFIP International Conference on Dependable Systems and Networks (DSN'07)*, pages 92–101, 2007.

[14] M. Saks and F. Zaharoglou. Wait-free k-set agreement is impossible: The topology of public knowledge. *SIAM J. Comput.*, 29(5):1449–1483, 2000.

[15] N. Santoro and P. Widmayer. Time is not a healer. In *Proc. 6th Annual Symposium on Theor. Aspects of Computer Science (STACS'89)*, LNCS 349, pages 304–313, Paderborn, Germany, Feb. 1989. Springer-Verlag.

[16] N. Santoro and P. Widmayer. Agreement in synchronous networks with ubiquitous faults. *Theor. Comput. Sci.*, 384(2-3):232–249, 2007.